\documentclass[letterpaper,10 pt,conference]{ieeeconf}

\IEEEoverridecommandlockouts
\usepackage[linesnumbered,algoruled]{algorithm2e}
\usepackage{amsmath}
\usepackage{amsfonts}
\usepackage{graphicx}
\usepackage{caption}
\usepackage{subfig}
\usepackage{color}
\usepackage{url}
\usepackage[mathscr]{euscript}
\graphicspath{ {./images/} }

\title{\LARGE \bf Multi-Robot Coordinated Planning in Confined Environments\\ under Kinematic Constraints}
\author{Clayton Mangette and Pratap Tokekar\thanks{C. Mangette is with the Department of Electrical and Computer Engineering, Virginia Tech, U.S.A. {\tt\small \{mangettecj\}@vt.edu}}\thanks{P. Tokekar is with the Department of Computer Science at the University of Maryland, U.S.A. {\tt\small \{tokekar\}@umd.edu}}}
\begin{document}
\maketitle
\thispagestyle{empty}
\pagestyle{empty}
\begin{abstract}
    We investigate the problem of multi-robot coordinated planning in environments where the robots may have to operate in close proximity to each other. We seek computationally efficient planners that ensure safe paths and adherence to kinematic constraints. We extend the central planner dRRT* with our variant, fast-dRRT (fdRRT), with the intention being to use in tight environments that lead to a high degree of coupling between robots. Our algorithm is empirically shown to achieve the trade-off between computational time and solution quality, especially in tight environments. The software implementation is available online at \url{https://github.com/CMangette/Fast-dRRT}.
\end{abstract}
\section{Introduction}
\indent Computationally efficient multi-robot motion planning algorithms are highly sought after for their applications in industry. In a time when automotive manufacturers are quickly approaching the advent of self-driving cars, centralized motion planners in lieu of traditional traffic control structures open the possibility of increased traffic flow in busy urban environments, with studies in \cite{v2x} and \cite{trafficflow} supporting this. With an increase in automation in warehouses by companies like Amazon \cite{ackerman_2019}, efficient path planning of robots designed to move inventory in place of human workers has become another important use case. Beyond ground vehicles, traffic management of Unmanned Aerial Vehicles (UAVs) is identified by NASA as an important area of research to ensure safe integration of aerial drones into the air space \cite{nasa}.
\\ \indent In each of the aforementioned applications, the algorithms used must be robust to planning in tight, confined environments while still ensuring that robots do not collide with one another. In the case of automated driving, urban traffic structures such as intersections and highway merging ramps constrain vehicles to a narrow set of paths. Similarly, warehouses limit robot paths due to shelving and storage units occupying the space. While not subject to high clutter, high volume air traffic can artificially restrict paths for UAVs. 
\\ \indent The planning algorithms available for such problems can be classified as centralized or decoupled. Centralized algorithms plan in the joint space of all robots whereas decoupled approaches only consider the space for each individual robot \cite{coupled_planners}. Decoupling interactions between robots that don't directly interact can simplify the original planning problem into a number of single-robot motion planning problems, making decoupled planners faster than centralized planners. However, this can compromise completeness and allow inter-robot collisions \cite{coupled_planners}. Centralized planners, in comparison, can guarantee collision-free motions and completeness, but at the cost of solution time and scale-ability. If a decoupled planner considers a space of dimension $\mathbb{R}^N$ for $d$ robots, then a centralized algorithm plans over a joint space $\mathbb{R}^{N^d}$. 
\\ \indent For our targeted applications, safety is of the utmost importance, so a centralized algorithm is better suited than a decoupled algorithm. Furthermore, centralized frameworks already exists in each use case. The intersection manager in \cite{IM_Centralized} is a hypothetical replacement to traffic lights that can control when autonomous vehicles enter an intersection via Vehicle-to-Infrastructure (V2I) communications. A task allocation and path planning system in \cite{amazon_robotics} demonstrates how to automate warehouse stock movement with kiva robots. The Unmanned Aerial System (UAS) Traffic Management (UTM) in development uses a centralized service supplier to manage requests and conflicts between UAVs operating within the same space \cite{nasa}. 
\begin{figure}[ht]
    \centering
    \includegraphics[width= 0.49\textwidth]{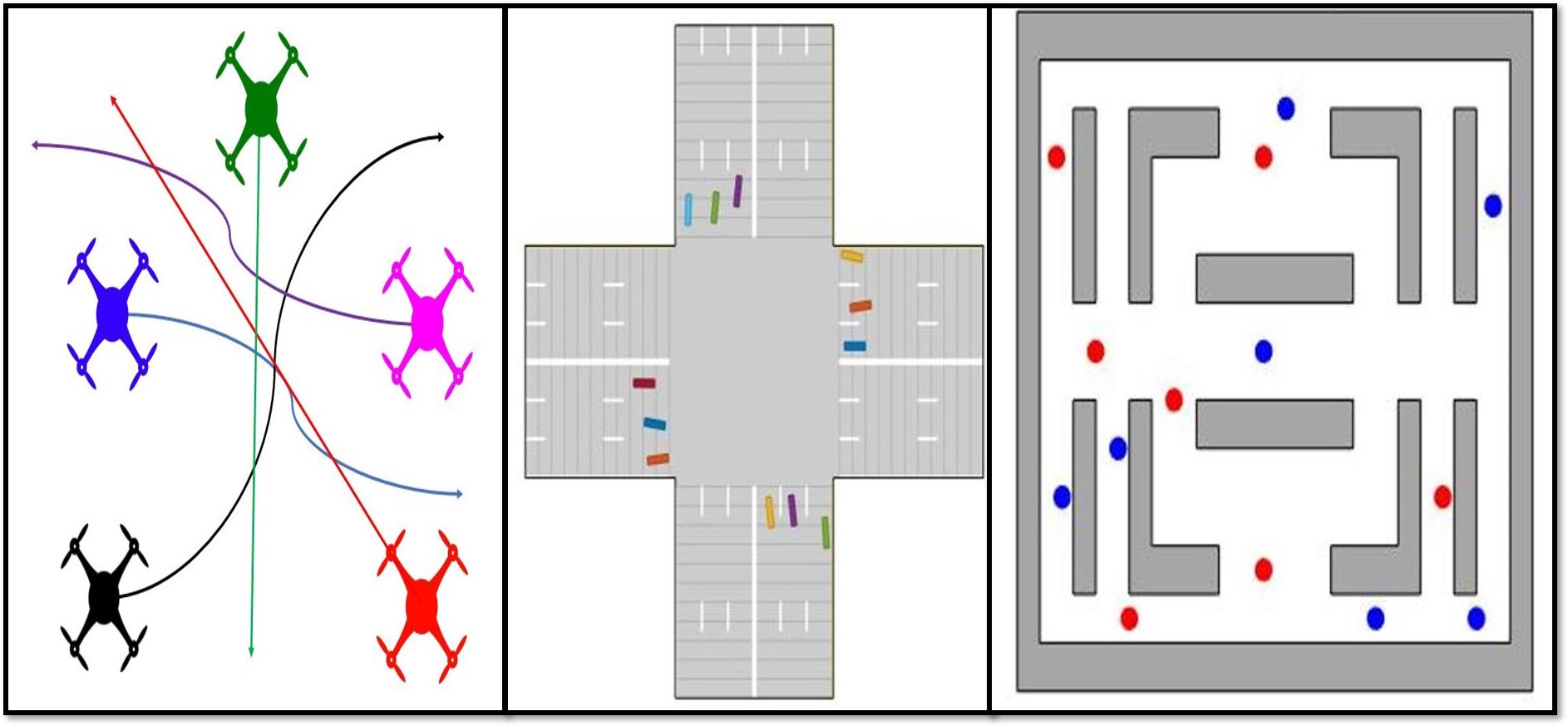}
    \caption{Targeted use cases: UAV coordination (left), traffic intersections (center), and warehouse motion planning (right).}
\end{figure}
\\ \indent The main challenge in centralized planning is doing so in a time-efficient manner. A secondary challenge is extending planning to robots with kinematic constraints, which complicates local path construction. This paper attempts solving both of these concerns by designing a framework for centralized planning in spaces with tight corridors with multiple kinematically constrained robots.
\\ \indent State-of-the-art planners have progressed towards algorithms that increase efficiency while preserving completeness. Recognizing the shortcomings of previous algorithms that rely on explicit computation of the composite planning space, discrete RRT (dRRT) \cite{drrt} and its optimal variant dRRT* \cite{drrt_star} improve computational efficiency by offloading computations to offline tasks when possible and relying on implicit representations on the planning space. These algorithms do not encode steering constraints, but provide a general framework for fast multi-robot planning.
\\ \indent This paper presents a variant to dRRT / dRRT* , which we call fast-dRRT (fdRRT), that returns fast but sub-optimal trajectories in tight environments requiring significant coordination between robots. We also extend these planners to account for robots with kinematic constraints. dRRT* and our algorithm are tested across multiple environments (Figure 1) and demonstrate fdRRT's increased computational efficiency in confined spaces. 
\section{Related Work}
\indent Motion planning has been studied as one of the fundamental problems in robotics. In the standard planning framework, a robot within a work space begins with a starting point $s$ and goal point $t$, and the solution to the planning problem is to find a collision-free path connecting $s$ and $t$. Grid-based methods such as Djikstra's algorithm \cite{Dijkstra1959} and A* \cite{A_star} were developed as a means of finding shortest paths between vertices on a graph. Sampling-based motion planners became popular for their adaptability to different kinematic models and low cost by sampling points instead of searching exhaustively over the work space. A detailed review of sampled-based planning is provided in \cite{sbmp_survey}. 
\\ \indent Extending motion planning to the multi-robot domain has been challenging due an increase in search space size and has led to a variety of approaches. Strategies are categorized in \cite{multi_survey} to use cell decomposition, potential field navigation, roadmaps to plan efficient paths. 
Cell decomposition methods to path planning rely on discrete maps of the planning space to determine optimal paths. A sequential process in \cite{d_star_multi} splits the problem into local path planning using D* and coordination between robots to avoid entering collision regions simultaneously. Instead of handling spatial and velocity planning separately,  Wagner and Choset developed M* , a multi-robot analogue to A* that resolves local path collisions by coupling paths only when they are found to overlap \cite{M_star}. Although M* can plan paths for up to 100 robots, its performance suffers when high degrees of coupling between robots arise at choke points in the planning space. Yu and Lavalle optimize paths on a graph across various objectives and demonstrate the scaleability of their algorithm, but do not consider kinematic contraints in their models \cite{optimal_graph_mpp}. 
\\ \indent Roadmap strategies, in contrast, iteratively explore the work space instead of searching exhaustively.  Van den Berg et al. provide a general framework for planning in a roadmap a sequential path planner that determines a sequential ordering for each robot to execute its path \cite{sequential_planner}. It relies on a coupled motion planner for handling local connections between conflicting agents, so run time performance is dependent on the degree of coupling between robots. The coordinated path planner in 
\cite{Coordinated_PP} searches collision-free paths over an explicitly computed multi-robot work space, but is limited in scope due to the memory required to build an explicitly defined road map. Using the principle of sub-dimensional expansion, Wagner et al. designed sub-dimensional RRT (sRRT) and sub-dimensional PRM (sPRM) to plan paths for multiple robots with integrator dynamics \cite{s_rrt}, the latter using M* to query a multi-robot path. 
\\ \indent Solovey et al. also use sub-dimensional expansion in discrete RRT (dRRT) \cite{drrt}. The idea of dRRT is to build road maps $\mathbb{G} = (G_1 , G_2 , ... , G_N)$ of collision-free motions for each robot, and then use them to build a search tree $\mathbb{T} = (V,E)$ implicitely embedded in $\mathbb{G}$. dRRT draws samples from each road map and combines them into a composite sample, $Q_{rand} = (q_{1,rand},q_{2,rand},... q_{N,rand})$, to which $\mathbb{T}$ is extended towards by selecting a composite neighboring vertex $V_{new}$. Because $\mathbb{G}$ relies on pre-computed motions between configurations that have already been collision checked against environmental obstacles, dRRT can simply fetch the motions $E_i \in G_i$ and check if any inter-robot collisions occur, thus relieving the algorithm of significant computational burden. Collision-free composite motions are added as vertices $V$ to $\mathbb{T}$ until a goal is reached.
\\ \indent The optimal variant of dRRT, dRRT*, improves upon computation time further by carefully choosing neighbors to expand towards the goal state \cite{drrt_star}. In addition to $\mathbb{G}$ , a path heuristic, $\mathbb{H}$, is computed to identify configurations with short paths to $Q_f$. This improves both solution quality and computational efficiency, making dRRT* the one of the state-of-the-art algorithms in multi-robot planning. 
\\ \indent This paper presents a centralized planning strategy for kinematically constrained robots in tight environments. We demonstrate the feasibility of our kinematically constrained PRM algorithm in extending the pre-existing dRRT algorithm to the domain of planning under motion constraints. The central planning algorithm, which we call fast-dRRT (fdRRT), is designed to switch between randomly exploring the state space and driving greedily towards the goal state in a manner similar to dRRT*. The difference in our algorithm is how expansion failures due to collisions are adjudicated. Instead of reporting an expansion failure if no collision-free connection can be established to a new node, fdRRT forces a connection by commanding some robots to stay in their previous configurations while permitting others to move forward. In practice, this makes fdRRT faster than dRRT* in tight work spaces, but at the cost of solution quality. Unlike dRRT*, our algorithm makes no guarantee of minimal path length, thus imposing an trade-off between solution efficiency and quality when choosing between the two algorithms. Additionally, fdRRT's incorporation of kinematic constraints makes it a more flexible planner that can be used in different systems.
\section{Problem Formulation}
The input to our problem is the set of start and goal positions for $N$ robots. The two goals are to construct a local map for each robot encompassing feasible paths connecting a robot's local start and goal position, and to use these maps to construct trajectories for each robot that respect kinematic constraints and do not intersect other trajectories.
\\ \indent Formally, given a set of initial configurations, $Q_{init} = (q_{1,init}, ... , q_{N,init})$ and final configurations, $Q_{goal} = (q_{1,goal} , ... , q_{N,goal})$, we would like to find a set of trajectories $\Pi = (\pi_1 , ... , \pi_N), \Pi(0) = Q_{init} , \Pi(1) = Q_{goal}$, such that all trajectories in $\Pi$ are non-intersecting with obstacles and other robots. Time is not explicitly is part of the configuration space, but we assume that each instance of a configuration $Q \in \Pi$ is uniformly discretized. Each robot is kinematically constrained by the motion model 
\begin{equation}
[\dot{x},\dot{y},\dot{\theta},\dot{\kappa}]^T = [\cos(\theta), sin(\theta), \kappa, \sigma ]^T
\end{equation}
For simplicity, we assume that each vehicle can only move forward. The dynamics in (1) are an extension of Dubins' steering constraints \cite{Dubins_1957} that add curvature constraints. The sum of path lengths of the multi-robot trajectory is the cost metric chosen for evaluation.
\section {Algorithm Overview}
\begin{figure}
\centering
\includegraphics[width = 0.40\textwidth]{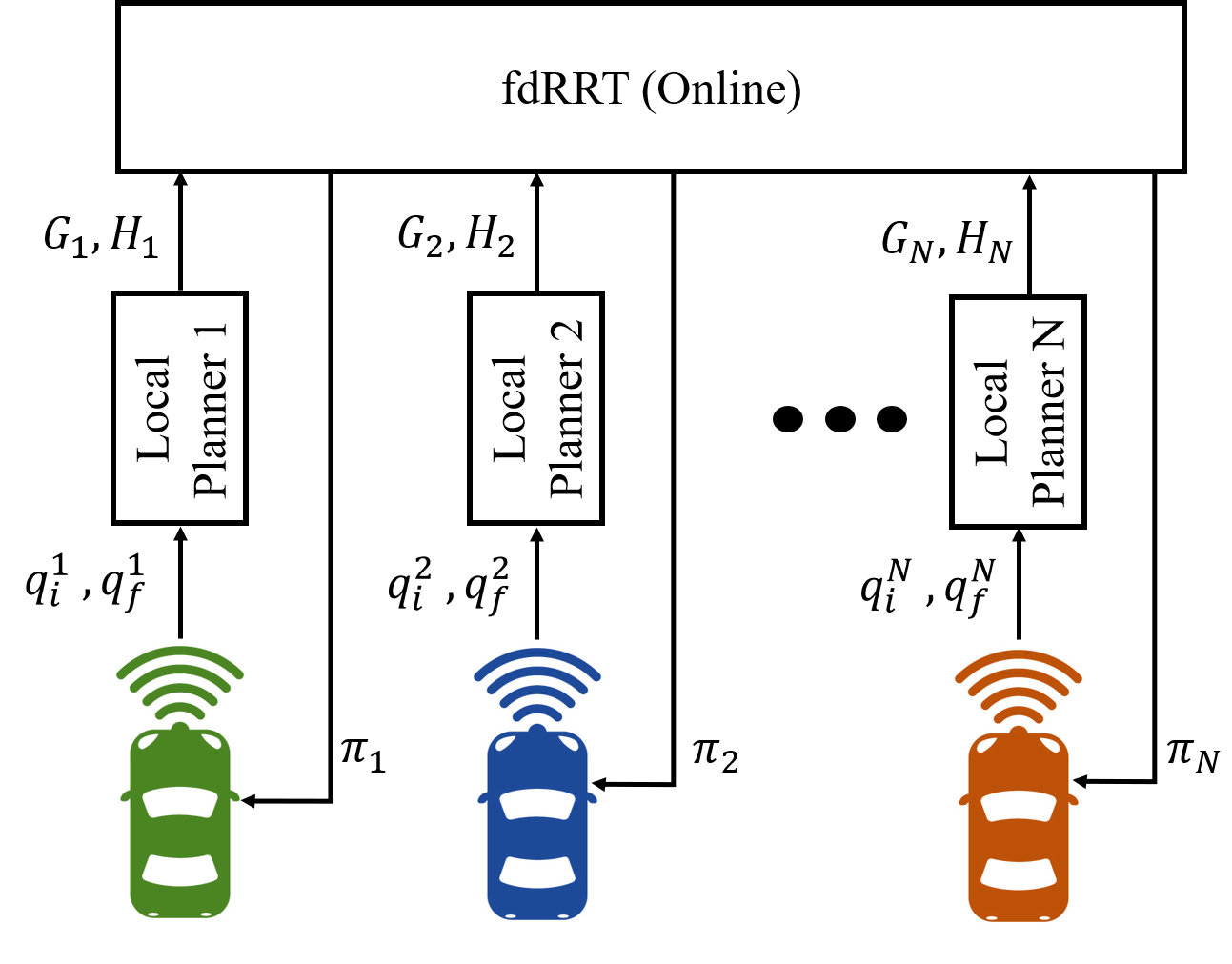}
\caption{The central planner returns collision-free path queries by referencing pre-computed roadmaps from a local planner.}
\label{block_diagram}
\end{figure}

Our system is illustrated in Figure \ref{block_diagram}. A local roadmap is constructed for each robot by the local planner that runs offline. The local roadmap is defined as a directed graph containing configurations within the robot's local configuration space and paths connecting configurations.
\\ \indent The central planner receives path queries in the form of initial and final configurations and local roadmaps from the robots entering the planning space. To avoid re-planning due to new requests, the central planner accepts requests until a deadline $T_{deadline}$ and relegate new requests to the next planning cycle. Given the local roadmaps and initial and final configurations of each robot, the central planner returns composite path $\Pi = (\pi_1 , \pi_2 , ... , \pi_R)$ that guarantees collision free trajectories between robots. Each local trajectory is sent to its corresponding robot as a list of time-parameterized waypoints $w_i(t) = [x_i(t) , y_i(t) ,\theta_i(t) , \kappa_i(t)]^T$ and connecting paths $\pi_i(s) = [x_i(s),y_i(s),\theta_i(s),\kappa_i(s)]^T$.
\\ \indent The local controller on each robot determines the speed profile to follow from $w_i(t)$ and the distance travelled between consecutive waypoints. $\pi_i(s)$ is re-parameterized to $\pi_i(t)$ from the distance traveled over time, which can be tracked by a local controller using a technique such as pure-pursuit or nonlinear Model Predictive Control (MPC) \cite{DBLP:journals/corr/PadenCYYF16}.
\subsection{Local Roadmaps}
Solovey et al. suggest using probabilistic roadmaps (PRMs) as approximations to local configuration spaces \cite{drrt}. The original PRM algorithm builds a roadmap as an undirected graph $G = (V,E)$, with each vertex $v \in V$ being a unique configuration and each edge $e(v_i , v_j) \in E$ a path in free space connecting two adjacent vertices $v_i$ and $v_j$ \cite{508439}. Configurations $q_{rand}$ are randomly sampled in the configuration space $C$ and connected to any vertices in $G$ within a connection distance $r$ , $ \{v \in V  | \texttt{dist(}q_{rand} , v) \leq r  \wedge e(q_{rand},v) \in C_{free}) \}$. Construction of $G$ continues for $N$ iterations, after which paths between configurations are found during a query.
\\ \indent This framework presents numerous challenges to adapting to a robot with kinematic constraints. Connections in \cite{508439} are line segments, which are sufficient under the assumption of single-integrator dynamics, but not for the dynamics in Equation (1). Numerical methods used in \cite{6161521} capture both kinematic and dynamic constraints to connect two configurations in a kinematically-constrained system, but are approximate solutions. Dubins’ paths adhere to kinematic constraints and yield minimal path length for car-like robots \cite{Dubins_1957}, but require sharp changes in steering curvature that are not achievable in a real system. Scheuer and Fraichard extend Dubins’ paths to continuous curvature paths using clothoids to transition between changes in curvature that, while less computationally tractable than Dubins’ paths, are a feasible connection method \cite{CC_Curves}. 
\\ \indent Additionally, $G$ in \cite{508439} is an undirected graph, implying that motions between connected vertices are bi-directional. Due to Dubins’ steering constraints and the non-holonomic constraints in (1), this is not necessarily true, and the existence of a collision-free path connecting two vertices $v_i$ to $v_j$ does not guarantee the reverse. To address this, Svestka and Overmars demonstrate that making $G$ a directed graph is sufficient to impose this restriction \cite{PPP_Car_Robots}.
\\ \indent Our local planner, kinematically-constrained PRM (KC-PRM), is similar to the Probabilistic Path Planner (PPP) in \cite{PPP_Car_Robots} with additional sampling and connection constraints to build a road map biased towards the optimal path that discriminates against unnecessary connections (Algorithm 1).
\begin{algorithm}
\SetAlgoLined
$\textrm{G} \leftarrow q_i$\;
\BlankLine
$\pi_{sample} \leftarrow \texttt{ReferencePath(}q_i , q_f\textrm{)}$\;
\BlankLine
\While{ $\texttt{Size(} \textrm{G}\texttt{)} < N$}{
\BlankLine
    $q_{rand} \leftarrow \texttt{RandomConfig(} \pi_{sample} \texttt{)}$\;
    \BlankLine
    \For{ $ v \in V$}
    {
        $\pi_{local} \leftarrow \texttt{Steer(}v , q_{rand} )$\;
        \BlankLine
        \If{$\texttt{IsReachable(} v , q_{rand}, \pi_{local}, r )$}
        {
            \If{$q_{rand} \notin \textrm{G}$}
            {
                $(\textrm{G}, v_{new} )\leftarrow \texttt{Insert(} q_{rand} \texttt{)}$\;
            }
            
            $\textrm{G} \leftarrow \texttt{Connect(}v,v_{new},\pi_{local}\texttt{)}$\;
        }
    }
}
\BlankLine
$\textrm{H} \leftarrow \texttt{CostToGoal(} \textrm{G} , q_f \texttt{)}$\;
\BlankLine
$\textrm{G} \leftarrow \texttt{PruneDeadNodes(} \textrm{G} , \textrm{H} \texttt{)}$\;
\Return{\textrm{G} , \textrm{H}}
\caption{\texttt{LocalPlanner(}$q_i , q_f , N , r$)}
\end{algorithm}
$G$ is initialized with an initial configuration $q_i$ (Line 1). A base path $\pi_{sample}$ is computed as the ideal path to follow from $q_i$ to $q_f$ and is used when sampling configurations (Line 2). $G$ expands to size $N$ by sampling random configurations $q_{rand}$, attempting connections to vertices (Lines 6 -- 7), and adding connections to $q_{rand}$ when attempts are successful (Lines 8 -- 11). Details are provide below.
\\ \indent $\texttt{RandomConfig}$: Random configurations are uniformly sampled along the sample path, $ q_{rand} \sim \mathscr{U} (\pi_{sample})$ with additive Gaussian noise, $\mathscr{N}(0,\sigma)$ to allow for variation in $q_{rand}$. The motivation behind sampling along $\pi_{sample}$ instead of the entire space is that one of the primary use cases is autonomous driving in urban environments. The space of locations that an autonomous vehicle can sample without violating traffic norms such as staying within one's own lane is confined to the center of the lane with some allowable deviation, so $q_{rand}$ is restricted appropriately.
\\ \indent $\texttt{Steer}$: Connections between adjacent vertices are attempted using the procedure described in \cite{CC_Curves}. Although more time consuming to compute than a Dubins' curve, this is an offline procedure, so efficiency is not a concern.
\\ \indent $\texttt{IsReachable}$: A configuration $q_2$ is defined to be reachable from $q_1$ if two conditions are met:
\begin{enumerate}
    \item the length of $\pi_{12} \leq r$, the path connecting $q_1$ to $q_2$, is within the connection radius.
    \item $q_2$ is in front of $q_1$. The planner in \cite{6161521} imposes a similar condition by checking if $q_2$ is in the half space of $q_1$. We check this condition by computing the normalized distance vector between $q_1$ and $q_2$ , $D = \frac{(x_2 - x_2 , y_2 - y_1)}{dist(q_1,q_2)}$, the tangent vector at $q_1$ , $T = (cos(\theta_1),sin(\theta_1))$, and check if the angle between these two vectors is less than 90 degrees.
\end{enumerate}
The purpose of this check is to only allow movements that would be feasible in traffic. Vehicle motions must move forward along the road in the direction of traffic, but this constraint isn't encoded into $\texttt{Steer}$. Thus, the reachability check enforces this behavior.
\\ \indent \texttt{CostToGoal}:The cost to go from each vertex in $G$ to $q_{goal}$ is stored in $H$ to be used as a heuristic in the central planner. Our implementation uses a breadth-first search.
\\ \indent \texttt{PruneDeadNodes}: Due to $G$ being a directed graph and the reachability constraints, some sampled nodes will not have a path to $q_f$. These "dead" nodes in $G$ are removed to avoid running into dead ends in the central planning stage.
\subsection{Central Planner}
\begin{algorithm}
\SetAlgoLined
\BlankLine
$\mathbb{T} \leftarrow Q_i$\;
$V_{last} \leftarrow Q_i$\;
\BlankLine
\While{$Q_f \notin \mathbb{T}$}
{
\BlankLine
$(\mathbb{T} , V_{last}) = \texttt{Expand(} \mathbb{T ,G ,H }, V_{last} , Q_f \texttt{)}$\;
    \BlankLine
    \If{$Q_f \in \mathbb{T}$}
    {
        $\Pi \leftarrow \texttt{FindPath(} \mathbb{T} , Q_f \texttt{)}$\;
        \Return{$\Pi$}
    }
}
\caption{\texttt{fdRRT}($Q_i, Q_f, \mathbb{G} , \mathbb{H}$)}
\end{algorithm}
\indent  The algorithm structure from dRRT* (Algorithm 1) is preserved with the initialization of $\mathbb{T}$ with $Q_i$ (Line 1). The algorithm then expands, while keeping track of the most recent expansion node $V_{last}$ to determine how it expands in the next iteration (Line 4). $\texttt{FindPath}$ queries $\mathbb{T}$ for a path to $Q_f$ and returns a composite path $\Pi$ if successful (Lines 5 -- 6). A notable difference in fdRRT is the omission of a local connector present in \cite{drrt} and \cite{drrt_star}, whose purpose is to solve the multi-robot coordination problem when sufficiently close to $Q_f$. We found this to be unnecessary in practice due to the structure of our environments. In the case of a traffic intersection, once all vehicles have passed through the physical intersection of the two roads, $\mathbb{T}$ tends to expand greedily towards $Q_f$. A similar subroutine is utilized in resolving path conflict by forcing some robots to hold their positions while others move forward.
\begin{algorithm}
\SetAlgoLined
\BlankLine

\uIf{ $V_{last} = \emptyset $}
{
    $Q_{rand} \leftarrow \texttt{RandomConfig(} \mathbb{G} \texttt{)}$\;
    $V_{near} \leftarrow \texttt{Nearest(} \texttt{T} , Q_{rand}\texttt{)}$\;
}
\Else
{

    $Q_{rand} \leftarrow Q_f$\;
    $V_{near} \leftarrow V_{last}$\;

}

$V_{new} \leftarrow \mathbb{I}_d(V_{near},\mathbb{G},\mathbb{H},Q_f)$\;
\BlankLine
$N \leftarrow \texttt{NeighborsInTree(} V_{new} , \mathbb{T} \texttt{)}$\;
\BlankLine
$( V_{best}^{free} , V_{best} ) \leftarrow \texttt{BestParent(}V_{new} , N \texttt{)}$\;
\BlankLine
\uIf{$V_{best}^{free} = \emptyset$}
{
    $V_H \leftarrow \texttt{ForceConnect(} V_{new},V_{best}\texttt{)}$\;
    \BlankLine
    \uIf{$V_H = \emptyset$}
    {
        \Return{$\emptyset$}
    }
    \Else
    {
        $\mathbb{T} \leftarrow \texttt{Connect(} V_{best} , V_H\texttt{)}$\;
        \Return{$V_H$}\;
    }
}
\Else
{
    $\mathbb{T} \leftarrow \texttt{Connect(} V_{best}^{free} , V_{new}\texttt{)}$\;
    \Return{$V_{new}$}\;
}

\caption{\texttt{Expand}($\mathbb{T}, \mathbb{G} , \mathbb{H} , V_{last} , Q_f$)}
\end{algorithm}
\\$\texttt{Expand}$: Expansion of $\mathbb{T}$ begins with selecting a node to expand from. If a vertex $V_{last}$ was added during the previous call, then a new expansion vertex $V_{new}$ is chosen by selecting a neighbor of $V_{last}$ (Lines 2--3). Otherwise, the closest neighbor $V_{near}$  of a random configuration $Q_{rand}$ is chosen (Lines 5 -- 6). The direction oracle subroutine selects an expansion node $V_{new}$ based on the success of the previous expansion (Line 8). If $Q_{rand} = Q_f$, $V_{new}$ is chosen as the tuple of individual vertices $v^i \in V$ that are neighbors to $v_{near}^i$ and have the lowest path cost to $q_f^i \in Q_f$, and is otherwise chosen as a tuple of randomly selected neighbors to $v_{near}^i$. We refer to \cite{drrt_star} for a detailed explanation.
\\ \indent All composite parents to $V_{new}$ that have already been added to $\mathbb{T}$ are expansion candidates to connect to $V_{new}$ (Line 9). Each candidate is evaluated base on whether the composite path between $N$ and $V_{new}$ results in a collision-free motion and the composite path cost. Our algorithm differs from \cite{drrt_star} when choosing the parent node to $V_{new}$, $V_{best}$. dRRT* chooses $V_{best}$ as the lowest cost vertex $V \in N$ that is also a collision-free motion. In our algorithm, the lowest cost collision-free node, $V_{best}^{free}$, and the lowest cost node $V_{best}$ are selected. If no such $V_{best}^{free}$ exists, the subroutine $\texttt{ForceConnect}$ attempts forcing $\mathbb{T}$ to expand by creating a new hybrid node, $V_H$, that restricts some individual nodes to hold their position at $v_{best}$, and allows others to move forward towards $v_{new}$. While forcing some vehicles to stop increases traffic delays for individual vehicles, $\texttt{ForceConnect}$ increases computational efficiency in practice by restricting random sampling to a last resort.
\begin{algorithm}
\SetAlgoLined
\BlankLine
$H \leftarrow \emptyset$\;
$L \leftarrow \emptyset$\;

$\Pi_{12} \leftarrow \texttt{LocalPaths(} V_1 , V_2\texttt{)}$\;

\For{$\pi_i \in \Pi_{12}$}
{
    \For{$\pi_j \in \Pi_{12} , i \neq j$}
    {
        $( H_i , L_i , A_i) \leftarrow \texttt{LocalPriority(}\pi_i, \pi_j\texttt{)}$\;
    }
}
$S \leftarrow \emptyset$\;
\For{$i = 1,2,..,N$}
{
    \uIf{$H_i = \emptyset \And A_i = \emptyset$}
    {
        $S \leftarrow S \cup i$\;
    }
    \ElseIf{$H_i = \emptyset \And A_i \neq \emptyset$}
    {
        \If{$\texttt{cost(}i\texttt{)} \leq \min(\texttt{cost(} j \in A_i\texttt{)})$}
        {
            $ S \leftarrow S \cup i$\;
        }
    }
}
$V_H \leftarrow \{ v_2^i | i \in S \} \cup \{ v_1^j | j \notin S \}$\;
\Return{$V_H$}\;
\caption{$\texttt{ForceConnect}(V_1 , V_2)$}
\end{algorithm}
\\$\texttt{ForceConnect}$: When forcing a connection between two composite nodes $V_1$ and $V_2$, the $i^{th}$ robot either holds its position at $v_1^i \in V_1$ or moves forward towards $v_2^i \in V_2$. Three sets are initialized for each robot $r_i \in R$: $H_i$, the set of robots with higher local priority than $r_i$, $L_i$, the set of robots with lower priority than $r_i$, and $A_i$, the set of robots that conflict with $r_i$ but have no local priority assigned. Each interaction is checked and $H,L, \textrm{ and } A$ are populated by $\texttt{LocalPriority}$. The local priority of $r_i$ with respect to $r_j$ is assigned according to the rules, which originate from the local connector logic in \cite{drrt} and \cite{sequential_planner}:
\begin{itemize}
    \item If $\pi_i(0)$ blocks $\pi_j$, then robot $i$ is given priority
    \item If $\pi_j(0)$ blocks the path of $\pi_i$, then robot $j$ is given priority
    \item If $\pi_i$ and $\pi_j$ do not overlap, then there is no interaction and a priority is not assigned
    \item Otherwise, the local priority can not be determined. This occurs when $\pi_i$ and $\pi_j$ overlap, but the starting positions of robots $i$ and $j$ do not block each other's paths. Either robot can be given priority, but the decision is deferred.
\end{itemize}
A solution set $S$ is then initialized to pick robots that should move forward (Line 9). Each robot is added to or rejected from $S$ based on its own $H_i , L_i , \textrm{and} A_i$ sets. For a robot $r_i$, if no other robots have a higher local priority and no robots have an undetermined priority, then $r_i$ is added to $S$. if any vehicles have a higher priority, then $r_i$ is rejected from $S$. If no robots have a higher priority, but some have undetermined priorities, then the cost of adding $r_i$ is assessed. In this context, the cost refers to number of vehicles that would be excluded from $S$ if $r_i$ was added to $S$. The cost of adding $r_i$ is compared to the cost of adding any of $r_j \in A_i$ and will be added to $S$ if the trade-off from adding $r_i$ is lower than the trade-off from adding any other member of $A_i$. After all robots are either added to or rejected from $S$, a hybrid node $V_H$ is constructed (Line 19).
\section{Simulations and Results}
\begin{figure*}[h]
   \subfloat[\label{tree_size}]{%
      \includegraphics[ width=0.33\textwidth,height=3.5cm]{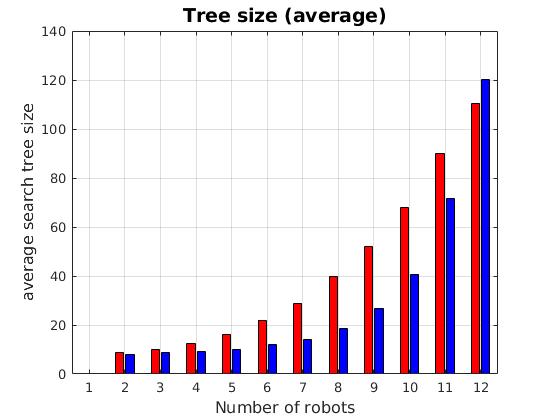}}
    \hspace{\fill}
   \subfloat[\label{solution_time}]{%
      \includegraphics[ width=0.33\textwidth,height=3.5cm]{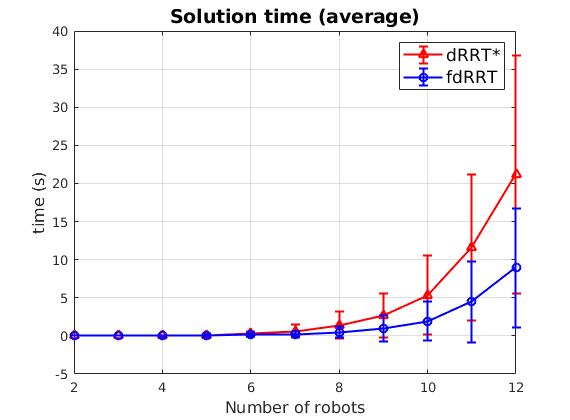}}
\hspace{\fill}
   \subfloat[\label{path_cost} ]{%
      \includegraphics[ width=0.33\textwidth,height=3.5cm]{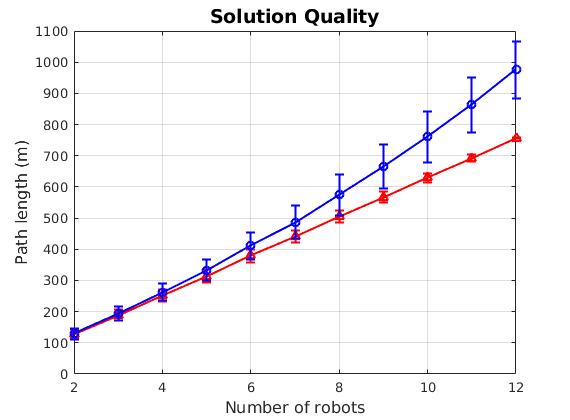}}\\
\caption{\label{intersection_results}Performance comparison in traffic intersection.}
\end{figure*}
\begin{figure*}[h]
    \subfloat[\label{tree_size_wh}]{%
      \includegraphics[ width=0.33\textwidth,height=3.5cm]{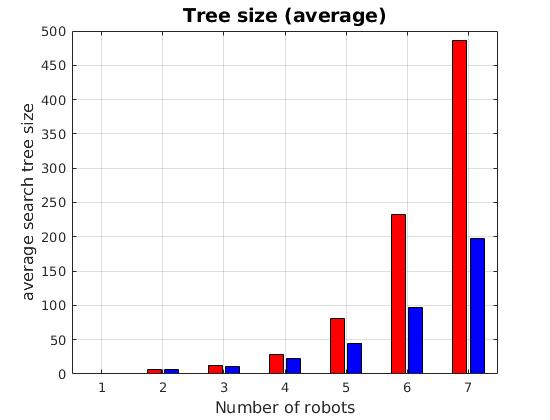}}
    \hspace{\fill}
   \subfloat[\label{solution_time_wh}]{%
      \includegraphics[ width=0.33\textwidth,height=3.5cm]{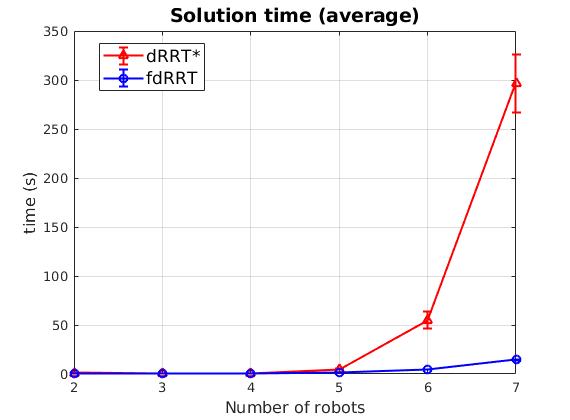}}
\hspace{\fill}
   \subfloat[\label{path_length_wh} ]{%
      \includegraphics[ width=0.33\textwidth,height=3.5cm]{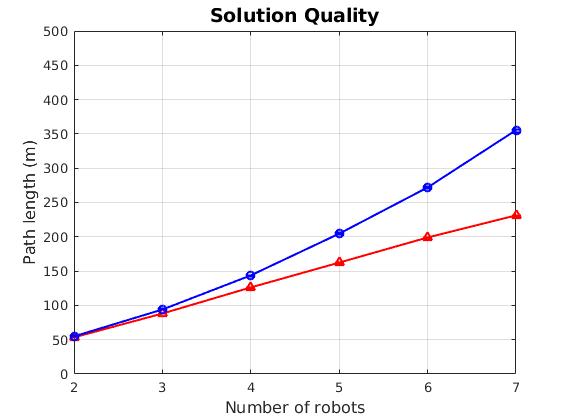}}\\
\caption{\label{wh_results}Performance comparison in a warehouse space.}
\end{figure*}
\begin{figure*}[h]
    \subfloat[\label{tree_size_uav}]{
      \includegraphics[ width=0.33\textwidth,height=3.5cm]{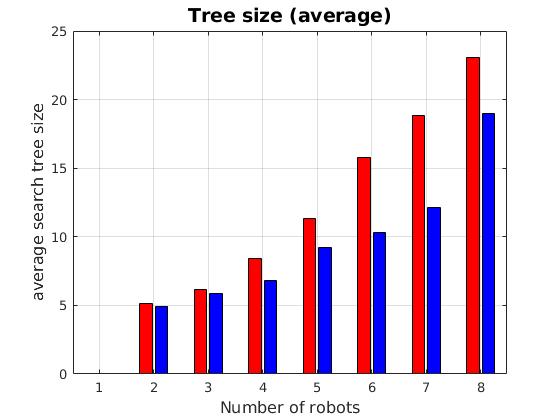}}
    \hspace{\fill}
   \subfloat[\label{solution_time_uav}]{%
      \includegraphics[ width=0.33\textwidth,height=3.5cm]{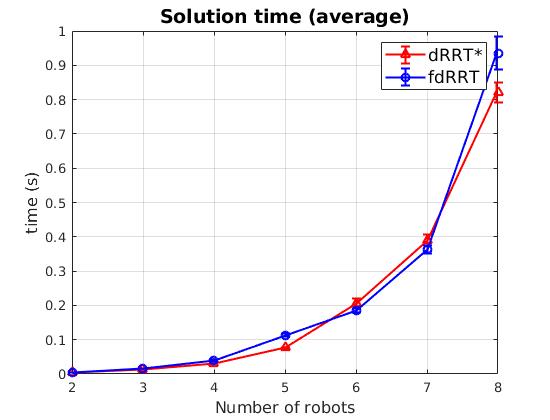}}
\hspace{\fill}
   \subfloat[\label{path_length_uav} ]{%
      \includegraphics[ width=0.33\textwidth,height=3.5cm]{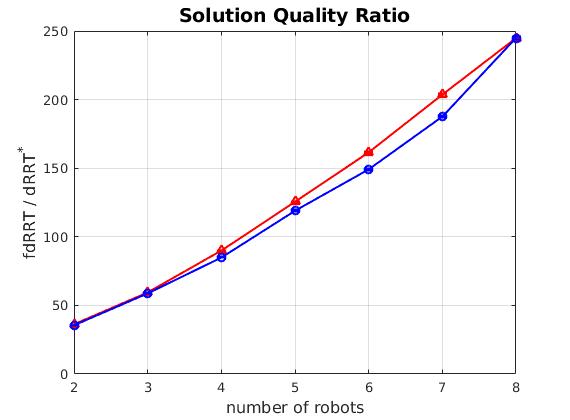}}\\
\caption{\label{wh_results}Performance comparison in UAV environment.}
\end{figure*}
Our algorithm was implemented and tested in MATLAB\@. Three environments are considered for validation: a three-lane traffic intersection, a cluttered rectangular space akin to a warehouse, and a crowded space of UAVs (Figure 1). We assume each vehicle in the first environment is rectangular with length $l = 3.6 m$ and $w = 1.6 m$, while the robots in the second and third environments disks with radius $r = 0.4 m \textrm{ and } 0.2m$, respectively. 1000 test cases were run for each combination of vehicles in each environment. Average search tree size, solution time, and path lengths are evaluation metrics to illustrate the trade-offs between dRRT* and fdRRT (Figures \ref{intersection_results} -- \ref{wh_results}).
\\ \indent From the test results, fdRRT performs better than dRRT* in computation efficiency in the intersection and warehouse spaces. In test cases with maximum traffic, fdRRT returned solutions 57\% faster in the traffic intersection and around 2000\% faster in the warehouse. However, dRRT* is 12\% faster in the UAV environment. This may be due to the lack of clutter within the UAV space, and thus reduced number of choke points. Under these conditions, the added computational time in fdRRT when forcing connections may degrade performance.  
\\ \indent Solution quality metrics show the opposite trend. As more robots are added to each environment, the path quality in fdRRT degrades, with paths being 22\% and 54\% longer in the intersection and warehouse spaces, respectively. Paths in the UAV space are nearly identical with a 0.2\% discrepancy.
\\ \indent The trends in solution times and path lengths across the scenarios can be attributed the amount of clutter in each space. The warehouse space has more obstacles distributed across its environment, and thus more possible choke points and corridors, the traffic intersection funnels all vehicles into a single, albeit large, choke point. The UAV space, in contrast, has no obstacles and thus allows the most movement. We conclude that there's a trade-off between the two algorithms; fdRRT will generally return trajectories faster, but dRRT* will have lower cost solutions.
\section{Conclusion}
We have developed a PRM planner for car-like robots to create trajectories that adhere to traffic standards, making it well suited for motion planning on roadway environments. That planning strategy was used in a central planning algorithm based on the previously published dRRT / dRRT* algorithm. Our implementation has demonstrated its advantage in computational time over dRRT* when planning in confined environments, at the cost of solution quality. 
\\ \indent The results from this study are promising, but several challenges remain. Testing the feasibility of fdRRT in a real system is one goal we would like to reach. We also plan to explore extending the planner to incorporate vehicle dynamics in addition to vehicle kinematics. In its current form, we only consider sampling configurations $q \in (x,y,\theta , \kappa)$ and ignore constraints on vehicle speed and acceleration. Adding constraints on vehicle dynamics makes connecting between configurations more difficult, but carries the benefit of ensuring that all paths are feasible for robots with both kinematic and dynamic constraints.
\bibliographystyle{IEEEannot}
\bibliography{Auto_dRRT_bib.bib}
\end{document}